\documentclass[twoside,twocolumn,9pt]{article}
\usepackage{extsizes}
\usepackage[super,sort&compress,comma]{natbib} 
\usepackage[version=3]{mhchem}
\usepackage[left=1.5cm, right=1.5cm, top=1.785cm, bottom=2.0cm]{geometry}
\usepackage{balance}
\usepackage{times,mathptmx}
\usepackage{sectsty}
\usepackage{graphicx} 
\usepackage{lastpage}
\usepackage[format=plain,justification=justified,singlelinecheck=false,font={stretch=1.125,small,sf},labelfont=bf,labelsep=space]{caption}
\usepackage{float}
\usepackage{fancyhdr}
\usepackage{fnpos}
\usepackage[english]{babel}
\usepackage{array}
\usepackage{droidsans}
\usepackage{charter}
\usepackage[T1]{fontenc}
\usepackage[usenames,dvipsnames]{xcolor}
\usepackage{setspace}
\usepackage[compact]{titlesec}
\usepackage{bm}
\usepackage{amsfonts}

\usepackage{dutchcal}

\usepackage{color, soul}

\definecolor{cream}{RGB}{222,217,201}

\begin{document}

\pagestyle{fancy}
\thispagestyle{plain}
\fancypagestyle{plain}{

\renewcommand{\headrulewidth}{0pt}
}

\makeFNbottom
\makeatletter
\renewcommand\LARGE{\@setfontsize\LARGE{15pt}{17}}
\renewcommand\Large{\@setfontsize\Large{12pt}{14}}
\renewcommand\large{\@setfontsize\large{10pt}{12}}
\renewcommand\footnotesize{\@setfontsize\footnotesize{7pt}{10}}
\makeatother

\renewcommand{\thefootnote}{\fnsymbol{footnote}}
\renewcommand\footnoterule{\vspace*{1pt}%
\color{cream}\hrule width 3.5in height 0.4pt \color{black}\vspace*{5pt}} 
\setcounter{secnumdepth}{5}

\makeatletter 
\renewcommand\@biblabel[1]{#1}            
\renewcommand\@makefntext[1]%
{\noindent\makebox[0pt][r]{\@thefnmark\,}#1}
\makeatother 
\renewcommand{\figurename}{\small{Fig.}~}
\sectionfont{\sffamily\Large}
\subsectionfont{\normalsize}
\subsubsectionfont{\bf}
\setstretch{1.125} 
\setlength{\skip\footins}{0.8cm}
\setlength{\footnotesep}{0.25cm}
\setlength{\jot}{10pt}
\titlespacing*{\section}{0pt}{4pt}{4pt}
\titlespacing*{\subsection}{0pt}{15pt}{1pt}

\fancyfoot{}
\fancyfoot[RO]{\footnotesize{\sffamily{1--\pageref{LastPage} ~\textbar  \hspace{2pt}\thepage}}}
\fancyfoot[LE]{\footnotesize{\sffamily{\thepage~\textbar\hspace{3.45cm} 1--\pageref{LastPage}}}}
\fancyhead{}
\renewcommand{\headrulewidth}{0pt} 
\renewcommand{\footrulewidth}{0pt}
\setlength{\arrayrulewidth}{1pt}
\setlength{\columnsep}{6.5mm}
\setlength\bibsep{1pt}

\makeatletter 
\newlength{\figrulesep} 
\setlength{\figrulesep}{0.5\textfloatsep} 

\newcommand{\topfigrule}{\vspace*{-1pt}%
\noindent{\color{cream}\rule[-\figrulesep]{\columnwidth}{1.5pt}} }

\newcommand{\botfigrule}{\vspace*{-2pt}%
\noindent{\color{cream}\rule[\figrulesep]{\columnwidth}{1.5pt}} }

\newcommand{\dblfigrule}{\vspace*{-1pt}%
\noindent{\color{cream}\rule[-\figrulesep]{\textwidth}{1.5pt}} }

\makeatother

\twocolumn[
  \begin{@twocolumnfalse}
\vspace{3cm}
\sffamily
\begin{tabular}{m{4.5cm} p{13.5cm} }
& 
\noindent\LARGE{\textbf{Critical behavior of quorum-sensing active particles 
}} \\
\vspace{0.3cm} & \vspace{0.3cm} \\

 & \noindent\large{Nicoletta Gnan \textit{$^{a,b}$$^\dag$} and Claudio Maggi,\textit{$^{c,b}$$^\ddag$}}
 \\

&
\noindent\normalsize{
\newline
It is still a  debated issue whether all critical active particles belong to the same universality class. Here we numerically study the critical behavior of quorum sensing active particles that represents the archetypal model for interpreting motility-induced phase separation. Mean-field theory predicts that this model should undergo a full phase separation if particles slow-down enough when sensing the presence of their neighbours and that the coexistence line terminates in a critical point. By performing large-scale numerical simulations we confirm this scenario, locate the critical point and use finite-size scaling analysis to show that the static and dynamic critical exponents of this active system agree with the Ising universality class.
} \\

\end{tabular}

 \end{@twocolumnfalse} \vspace{0.6cm}
]

\renewcommand*\rmdefault{bch}\normalfont\upshape
\rmfamily
\section*{}
\vspace{-1cm}


\footnotetext{\textit{$^{a}$~CNR-ISC, Institute of Complex Systems, Roma, Italy}}
\footnotetext{\textit{$^{b}$~Dipartimento di Fisica, Universit\`{a} di Roma ``Sapienza'', I-00185, Roma, Italy}}
\footnotetext{\textit{$^{c}$~NANOTEC-CNR, Institute of Nanotechnology, Soft and Living Matter Laboratory - Piazzale A. Moro 2, I-00185, Roma, Italy}}
\footnotetext{\textit{$^\dag$ E-mail: nicoletta.gnan@cnr.it}}
\footnotetext{\textit{$^\ddag$ E-mail: claudio.maggi@cnr.it}}



\section*{Introduction}
Quorum sensing (QS) is a widely exploited  communication strategy among microorganisms which allows a single cell to \emph{sense} the concentration of the population and act accordingly via genes expression~\cite{MillerAnnualRev2001}. A famous example is that of the bacterium \textit{Aliivibrio fischeri}~\cite{Nealson1970} which is capable of emitting bioluminescence depending on the density of the colony. More recently, it has been shown that QS is implicated in several other biological processes, such as virulence~\cite{ZhuPNAS2002} or biofilm formation~\cite{Hammer2003} and that QS can be used by cells to regulate their motility~\cite{Daniels2004}.

In recent years, scientists have been questioning how to mimic biology by designing synthetic active particles able to display collective behaviours similar to QS. Such artificial self-propelled particles are typically phoretic colloids exploiting self-generated chemical\cite{Gomez-SolanoSciRep2017} or temperature gradients~\cite{PhysRevLett.105.268302} to achieve autonomous locomotion. Advances in the control of these active particles have allowed to create nano- and micro-robots  capable of accomplishing some particular tasks~\cite{maggi2016self} and also to realize active systems with controllable interactions. This is the case of light-activated Janus particles whose speed is controlled by the intensity of the impinging radiation. In these systems QS-type interaction is achieved thanks to a real-time particle-detection algorithm which calculates the density field from particles position and adjusts their speed accordingly via light-modulation~\cite{SpeckNatComm2018}. A similar technique has been employed to implement active particles capable to adjust their velocities depending on the direction of their peers~\cite{Lavergne2019}. In both cases it has been observed that particles may act collectively by forming dense clusters. 

These aggregation phenomena can be generally rationalized within the so called motility induced phase separation (MIPS) scenario that has attracted a deep interest in the active matter community~\cite{CatesAnnRev2015}. MIPS broadly indicates phase separation occurring in active particles which slow down when the local density is high. These active particles thus further accumulate in slow regions causing a positive feedback loop triggering condensation. It has been also shown that such a condensation process, at the mean-field level, is completely analogous to the standard gas-liquid phase separation. Moreover the MIPS framework has been used to explain the condensation observed in active particles interacting via purely repulsive interaction potentials. In such a systems inter-particle collisions are responsible for the dynamic slow-down and, in the modelling, it is assumed that the net effect of pairwise interactions can be mapped into a density-dependent particle speed that decreases when local density increases, as in QS models. 
More recently, a formalism including both QS and pairwise interactions, has been proposed~\cite{SolonPRE2018} to describe MIPS. 

Besides the interest in studying fully phase-separated active particles, the attention of the community has been recently turned to the study of active systems close to the ending point MIPS curve i.e. in the vicinity of the \textit{motility induced critical point}. However these studies have reached different conclusions about the universality classes of these systems. For instance, it has been shown that Active Ornstein-Uhlenbeck particles (AOUPs) in two dimensions, interacting with pairwise repulsion, have static and dynamic critical exponents that are in agreement with the Ising universality class~\cite{maggi2021universality, MaggiCommPhys2022}. Differently, results for active Browinan disks are controversial: some in-lattice simulations point towards Ising~\cite{PhysRevLett.123.068002} while other in- and off-lattice simulations provide critical exponents deviating considerably from the Ising ones and suggest that some active systems may belong to a different universality class~\cite{siebert2018critical, dittrich2021critical}. It is important to mention that the evaluation of critical exponents in these off-equilibrium systems is highly challenging, since one cannot use the same tools routinely employed in the study of critical equilibrium systems such as grand-canonical simulations. Moreover it has been pointed out that these systems develop additional (non-universal) correlations~\cite{maggi2021universality}~ associated with the formation of hexatic microdomains~\cite{PhysRevLett.125.178004} and to clusters of particles with aligned velocities~\cite{caprini2020hidden}. Since these correlations play a role at intermediate time a length scales, one must use very large system sizes for observing the scaling regime in critical active systems.

All these results have been obtained for model systems with pairwise repulsive interactions, while no study has focused so far on critical QS active particles. In this work we fill this gap by investigating the critical behavior of a minimal model of AOUPs with 
\textit{short-ranged} QS interactions in two dimensions (${d=2}$). 
AOUPs possibly represent the simplest model for active particles since, as detailed below, the active propulsion force is produced by a linear stochastic process whose analytical properties have allowed to derive numerous theoretical results~\cite{Marconi3,Fodor16,Marconi17,dal2019linear}.
Moreover it has been recently shown that QS AOUPs undergo MIPS if the model's control parameters are varied appropriately~\cite{martin2021statistical}.
We first study the model at the mean field-level showing that the system phase separates when the parameter controlling the particles' slowing down is varied. Moreover we show that the spinodal line terminates in a critical point. By performing computer simulations in $d=2$ we find that the mean-field theory is in \textit{qualitative} agreement with the numerical results. By performing a detailed finite-size-scaling analysis we locate the critical point and estimate independently from each other the static critical exponents $\nu$, $\gamma$, $\beta$ and $\eta$ finding that these substantially agree with the Ising exponents. Lastly, we characterize the critical dynamics by studying the density fluctuations at large scales and their relaxation frequencies. We find that also the dynamic exponent $z$ is compatible with the Ising value.

\section*{Theory}

Here we show that our minimal model of QS AOUPs has a mean-field phase diagram characterized by a spinodal line ending in critical-point. As we will see in the following, the control parameter that can be varied to induce the transition is the speed variation that a particle undergoes when it ``senses'' a local density change. 

We start by considering one single AOUP having position $\mathbf{x}$ and subjected to a space-dependent scalar speed-field 
$\mathcal{v}(\mathbf{x})$ (here $\mathbf{x}={x_1,...,x_d}$ in a $d$-dimensional space). The equations of motion of such an AOUP are given by:

\begin{eqnarray}
\label{vx}
\dot{\mathbf{x}} &=& \mathcal{v}(\mathbf{x}) \, 
\boldsymbol{\psi}\\
\label{vx2}
\tau \, \dot{\boldsymbol{\psi}} &=& -\boldsymbol{\psi} + \boldsymbol{\eta}
\end{eqnarray}

\noindent where $\tau$ is the persistence time of the active force $\boldsymbol{\psi}$ and $\boldsymbol{\eta}$ is a standard white noise source, i.e. $\langle \eta_\alpha(t) \rangle=0$ and ${\langle \eta_\alpha(t)\, \eta_\beta(s) \rangle = 2 D \, \delta_{\alpha\beta} \, \delta(t-s)  }$, where the greek indices indicate the Cartesian components. 
The stationary probability density $p(\mathbf{x})$ for the process (\ref{vx})-(\ref{vx2}) reduces to the standard $p(\mathbf{x})$ of a (Stratonovich) process driven by multiplicative white noise~\cite{risken1996fokker,gardiner1985handbook,Hanggi95}: 

\begin{equation}
\label{pv}
p(\mathbf{x}) = \frac{Z^{-1}}{\mathcal{v}(\mathbf{x})}
\end{equation}

\noindent where $Z=\int d\mathbf{x}\,[\mathcal{v}(\mathbf{x})]^{-1}$. We now assume that the speed of the probe particle in $\mathbf{x}$ is determined by the positions $\mathbf{x}_j$ of the other $N$ particles in the system, i.e. we set ${\mathcal{v}(\mathbf{x})=\mathcal{v}(\mathbf{x},\mathbf{x}_j)}$
in Eq.~in (\ref{pv}). Note that, in this approximation, the degrees of freedom $\mathbf{x}_j$ are considered as if they were \textit{frozen} parameters. We further assume that the individual $\mathbf{x}_j$ contribute additively to the speed in $\mathbf{x}$ via some function $g$:

\begin{equation}
\label{vg}
\mathcal{v}(\mathbf{x}) = \mathcal{v}\left(\sum_j g (\mathbf{x},\mathbf{x}_j)\right)
\end{equation}\\

\noindent 
The function $g$ is chosen to depend only on the distance between the particle in $\mathbf{x}$ and its neighbour in $\mathbf{x}_j$, i.e. ${g(\mathbf{x},\mathbf{x}_j)=g (|\mathbf{x}-\mathbf{x}_j|)}$.
If, for example, $d=2$ and

\begin{equation}
\label{gTheta}
g (|\mathbf{x}-\mathbf{x}_j|) = 
\Theta(R-|\mathbf{x}-\mathbf{x}_j|)/(\pi R^2)
\end{equation} 

\noindent (with $\Theta$ being the Heaviside step-function) then the speed of the particle in $\mathbf{x}$ is determined by the local density $g$ measured by the particle in a disk of radius $R$. Note that the form of $v$ does not need to be specified at this point. By rewriting (\ref{vg}) in terms of delta functions and using the mean-field approximation we have

\begin{equation}
\label{vg2}
\mathcal{v}(\mathbf{x}) = \mathcal{v}\left(\int 
d\mathbf{x}' \, g(|\mathbf{x}-\mathbf{x}'|) \, \rho(\mathbf{x}') \right)
\end{equation}

\noindent Note that $\mathcal{v}(\mathbf{x})=\mathcal{v}[\rho(\mathbf{x})]$ can  now be seen as a functional of the density field 
${\rho(\mathbf{x})=\langle\sum_j\delta(\mathbf{x}-\mathbf{x}_j)\rangle}$.
Moreover the gradient expansion of (\ref{vg2}) is~\cite{vanKampen}:

\begin{equation} 
\label{exp02}
\int d
\mathbf{x}' g(|\mathbf{x}-\mathbf{x}'|) \, \rho(\mathbf{x}')
\approx 
g_0 \, \rho(\mathbf{x}) + g_2 \, \nabla^2\rho(\mathbf{x})
\end{equation}

\noindent 
where $
g_0 = \int d\mathbf{x} \,g(|\mathbf{x}|)$ and $g_2 =\int d\mathbf{x} |\mathbf{x}|^2\,g(|\mathbf{x}|)/2$ which are constants independent on $\mathbf{x}$.
From now on we assume for simplicity that $g_0=1$ (as in the example (\ref{gTheta})). Since, for identical particles $N p(\mathbf{x})= N \langle \delta(
\mathbf{x}-\mathbf{x}_j) \rangle = \rho(\mathbf{x})$, by taking the log of Eq.~(\ref{pv}), we obtain the self-consistency equation:

\begin{equation}
\label{seqv1}
\ln \rho(
    \mathbf{x}) + \ln\left[ \mathcal{v} \left( 
\rho(\mathbf{x}) + g_2 \, \nabla^2\rho(\mathbf{x})
\right) \right] = \mu
\end{equation}

\noindent where $\mu=\ln(N/Z)$ is an effective chemical potential which can be thought as a Lagrange multiplier adjusting the total number of particles $N$ in the system~\cite{vanKampen}. By using (\ref{exp02}) in (\ref{seqv1}) and further expanding in gradients we have

\begin{equation} 
\label{seqv2}
\ln \rho(\mathbf{x}) + \ln \mathcal{v} \left( \rho(\mathbf{x}) \right)  + g_2 \frac{\mathcal{v}'(\rho(\mathbf{x}))}
{\mathcal{v}(\rho(\mathbf{x}))}\nabla^2\rho(\mathbf{x})
=
\mu
\end{equation}

\noindent where $\mathcal{v}'(\rho)=\partial_\rho \mathcal{v}(\rho)$. 

Assuming an homogeneous density field $\rho(\mathbf{x})=\rho$ Eq.~(\ref{seqv2}) reduces to the fundamental equation describing MIPS at the mean-field level~\cite{Tailleur08,cates2015motility}

\begin{equation} 
\label{seqMF}
\ln \rho + \ln \mathcal{v} \left( \rho \right) 
=
\mu
\end{equation}

For the choice of $\mathcal{v}(\rho)$ we now require that (\textit{i}) Eq.~(\ref{seqMF}) admits a positive solution ($\rho>0$) for any number of particles
${0<N<\infty}$ (i.e. for ${-\infty<\mu<\infty}$) and that (\textit{ii}) Eq.~(\ref{seqMF}) admits more than one positive solution, at least in some parameter range. A possible choice for $\mathcal{v}(\rho)$, which satisfies (\textit{i}) and (\textit{ii}), is:

\begin{equation}
\label{vofr}
\mathcal{v}(\rho) = \frac{\Delta \mathcal{v}}
{1+(\rho/\rho_0)^2}+\mathcal{v}_0
\end{equation}

\noindent 
where $\Delta \mathcal{v},\mathcal{v}_0$ and $\rho_0$ are positive parameters which control the response of the particle's speed to the density.
According to (\ref{vofr}) when a particle ``percieves'' a density much lower than the threshold density $\rho_0$ it moves fast (close to maximum speed $\mathcal{v}_0+\Delta \mathcal{v}$), while it moves almost at minimum speed ($\mathcal{v} \approx \mathcal{v}_0$) when $\rho \gg \rho_0$.
Property (\textit{i}) is clearly satisfied by (\ref{vofr}) since $\ln \rho$ ranges from $-\infty$ to $\infty$ (as $\rho$ varies between $0$ and $\infty$) while ${\ln \mathcal{v}_0<\ln \mathcal{v}(\rho)}<\ln(\Delta \mathcal{v}+\mathcal{v}_0)$ stays finite. Property (\textit{ii}) is fulfilled if $\partial_\rho\mu<0$, i.e. if $\mathcal{v}'(\rho)/\mathcal{v}(\rho)<-1/\rho$ (which is the condition for MIPS~\cite{Tailleur08}). Assuming that $\mathcal{v}_0$, $\rho_0$ are fixed in (\ref{vofr}), while $\Delta \mathcal{v}$ is the control parameter, the system undergoes spinodal decompostion when

\begin{equation}
\label{spin}
\Delta \mathcal{v} >\frac{\left(\rho ^2+\rho _0^2\right){}^2 \mathcal{v}_0}{\rho _0^2 \left(\rho ^2-\rho _0^2\right)}
\end{equation}

\begin{figure*}[t!]
 \centering
\includegraphics[width=1.0\textwidth]{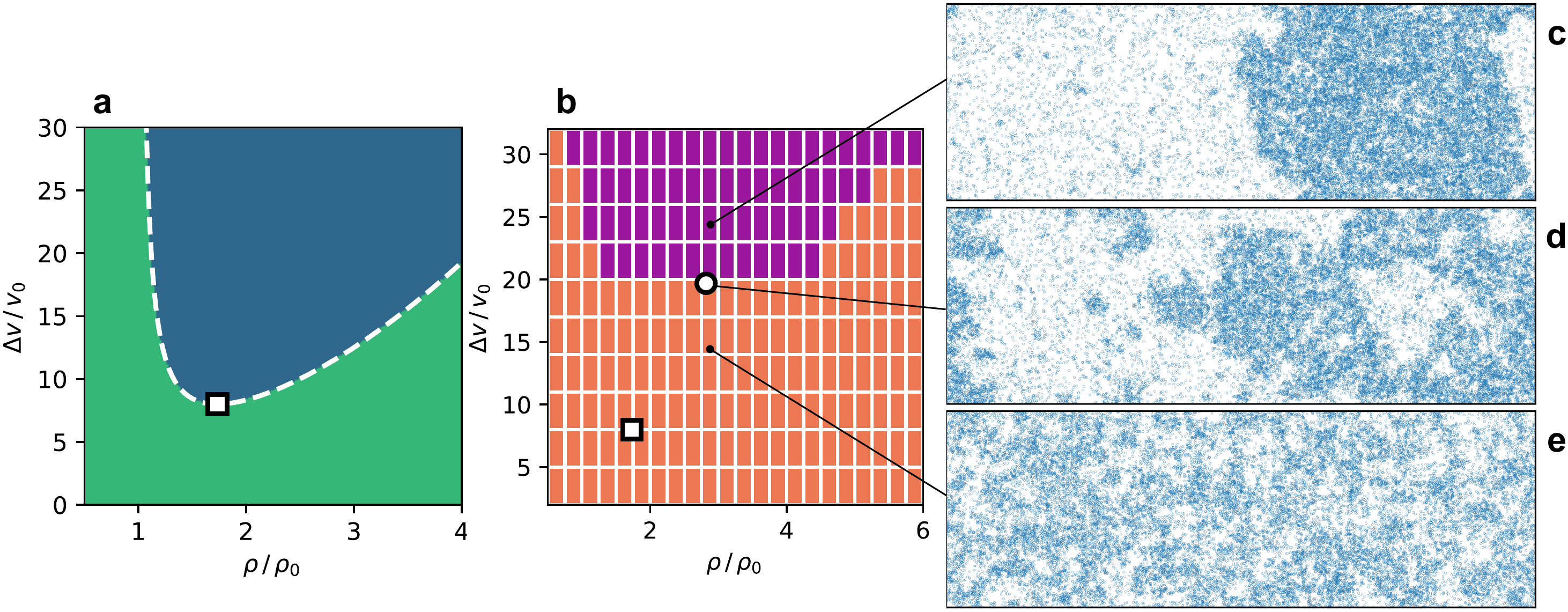}
\caption{
\label{sktch}
(\textbf{a}) Theoretical spinodal curve (dashed line) from the mean field model of QS active particles.
The system undergoes spinodal decomposition for large values of $\Delta \mathcal{v}$ and intermediate values of $\rho$ (blue area). The spinodal line terminates in a critical point (open square). (\textbf{b}) Coexistence region and critical point in AOUPs interacting via QS in $d=2$ (numerical simulations).
The system phase-separates for large values of $\Delta \mathcal{v}$ and intermediate values of $\rho$ (purple area).
The estimated critical point is represented by an open circle. The critical point of the mean-field model in (a) is reported here for comparison (open square). (\textbf{c})-(\textbf{e}) Snapshots of the largest system investigated ($N=30\times10^3$ particles) at three different $\Delta \mathcal{v}$-values indicated in (b) along the critical density: a phase separated (c), a near-critical (d)  and a homogeneous state point (e).}
\end{figure*}

\noindent Moreover, by solving
\begin{eqnarray}
\partial_\rho\mu|_{\rho=\rho_c,\Delta \mathcal{v}=\Delta \mathcal{v}_c}&=&0\\ \partial^2_{\rho^2}\mu|_{\rho=\rho_c,\Delta \mathcal{v}=\Delta \mathcal{v}_c}&=&0
\end{eqnarray}

\noindent we find that the spinodal line defined by (\ref{spin}) terminates in a critical point located at $\rho_c=3^{1/2}\rho_0$ and $\Delta \mathcal{v}_c=8\,\mathcal{v}_0$ as sketched in Fig.~\ref{sktch}(a). 

Interestingly such a schematic phase diagram suggests that when particles react moving \textit{faster} in low-density regions (i.e. when $\Delta \mathcal{v}$ is increased above $\Delta \mathcal{v}_c$) the system may phase separate, while particles always keep a non-zero minimum speed $\mathcal{v}_0$ also in the dense phase. 
Moreover an expansion of Eq.s~(\ref{seqMF}) around the critical point shows that the latter belongs to the (mean-field) Ising universality class in full analogy with the condensation of Van der Waals fluids~\cite{vanKampen}.
In the following sections we will compare this scenario with results of direct particle simulations in $d=2$.

\section*{Numerical Model and Methods}\label{sec:Simulation}

In the microscopic model we consider $N$ identical  point-like AOUPs (no excluded volume interactions) that move according to Eq.s~(\ref{vx}) and (\ref{vx2}). Each particle interacts with its neighbours first by measuring the local density via Eq.~(\ref{vg}) and then by adjusting its speed via Eq.~(\ref{vofr}).
For simplicity we have fixed all dynamic parameters (${D=\tau=\mathcal{v}_0=\rho_0=1}$ and $R=1.3$) but $\Delta \mathcal{v}$ which we will vary to induce the phase separation and hence, together with density, it represents a control parameter of the system. Particles move in a (two-dimensional) rectangular box with periodic boundary conditions and with sides $L_x=3\times L_y$. To apply the finite-size scaling analysis for the study of the critical point, we simulate different system sizes with ${N = (3.5,7.5, 15, 30)\times 10^3}$. The equations of motion have been integrated numerically using the Euler scheme with a time step ${\Delta t=10^{-3}}$ up to ${N_t =3\times 10^8}$ time steps for the largest system size. To reach the stationary state a preliminary run has been performed for a comparable amount of steps, starting from a configuration in which particles are randomly distributed. As in~\cite{maggi2021universality} we have checked that the auto-correlation function of density fluctuations relaxes for all system sizes within the simulation time-window. Moreover, for each state point investigated, the whole procedure has been repeated five times starting from independent configurations.

To characterize the critical behavior of our model, we will rely on the finite size scaling ansatz according to which any observable $\mathcal{O}$ can be rewritten in terms of a dimensionless scaling function $G_\mathcal{O}(L /\xi)$:

\begin{equation}
\mathcal{O} = L^{\frac{\zeta_\mathcal{O}}{\nu}}\, G_\mathcal{O}(L /\xi)    
\label{fss}
\end{equation}

\noindent where $\zeta_\mathcal{O}$ is the critical exponent associated with the observable $\mathcal{O}$ and $\nu$ is the exponent characterizing the divergence of the correlation length $\xi$ as the control parameter gets close to its critical value. As suggested by the mean-field theory described above, the speed change parameter  $\Delta \mathcal{v}$ can be used as a control parameter to tune the system at criticality and thus we assume  $\xi \sim |\Delta \mathcal{v}-\Delta \mathcal{v}_c|^{-\nu}$.  Note that the scaling ansatz~(\ref{fss}) implies that, when the quantity ${L^{-\zeta_\mathcal{O}/\nu} \, \mathcal{O}}$ is plotted as a function of ${L^{1/\nu} (\Delta \mathcal{v}-\Delta \mathcal{v}_c)}$,  for different system sizes, all data should collapse onto the scaling function.

We are interested in measuring observables~$\mathcal{O}$ related to density fluctuations both for locating the critical point and estimating the critical exponents. These fluctuations are calculated using an improved version of the ``block-density-distribution method'' initially proposed by Binder and coworkers~\cite{binder1981finite,rovere1988block,rovere1990gas,rovere1993simulation}. This technique is an alternative to grand-canonical simulations (where the particle density fluctuate) to characterize density fluctuations in the canonical ensemble (where the overall density is fixed). The main idea is to divide the simulation box into small sub-boxes of linear size~$L$ that exhibit, to a good approximation, the same density fluctuations of the grand-canonical ensemble. In each sub-box one measures the particles number density
$\rho=N_b/L^2$, where $N_b$ is the number of particles found in a sub-box, and compute the density variance
${\langle\Delta\rho^2\rangle=\langle(N_b/L^2)-(\langle N_b/L^2\rangle)^2\rangle}$ and its fourth moment ${\langle\Delta\rho^4\rangle=\langle(N_b/L^2)-(\langle N_b/L^2\rangle)^4\rangle}$. From these quantities one then computes the ``Binder cumulant'' ${\mathcal{B}=\langle\Delta\rho^2\rangle^2/\langle\Delta\rho^4\rangle}$ for which one expects that the ansatz (\ref{fss}) reduces to the simple form $\mathcal{B}=G_\mathcal{B}(L/\xi)$ and thus, at the critical point, $\mathcal{B}$ is size independent since ${\mathcal{B}(\xi\rightarrow \infty)=\mathrm{const}}$. This suggests that one could locate the critical point by finding the value of the control parameter at which the $\mathcal{B}$ curves, for different sizes, cross. Unfortunately it has been shown that this procedure fails (the $\mathcal{B}$-curves do not cross) even for the equilibrium two-dimensional lattice gas in the canonical ensemble (which is equivalent to the Ising model with conserved magnetization)~\cite{rovere1993simulation, siebert2018critical}. The failure of this scheme can be attributed to the fact that the grand-canonical fluctuations cannot be well reproduced in the sub-boxes due to the large contribution coming from the interface between the two phases. 

To overcome this problem, a modified version of the method has been proposed in Ref.~\cite{siebert2018critical} and successfully applied both to critical equilibrium
systems and active systems~\cite{siebert2018critical,PhysRevLett.123.068002,maggi2021universality,dittrich2021critical}.
The basic idea of the modified method is that the system center of mass can be used to locate the dense phase excluding the interfaces from the analysis. More specifically for each particle configuration the $x$-coordinate of the systems' center of mass is always shifted to $x=3\,L_x/4$ and density fluctuations are then evaluated far from the interface, by selecting only four squared sub-boxes of size $L=L_y/2$: two sub-boxes centered around $x=3\,L_x/4$ (dense phase) and the other two in $x=L_x/4$ (dilute phase). The same procedure is then repeated for different system sizes to find the crossing of the cumulants $\mathcal{B}$ and thus locate the critical point. It has been shown that this technique yields the correct critical point in $d=2$ in the case of equilibrium square-lattice gas~\cite{siebert2018critical} and the triangular-lattice gas~\cite{maggi2021universality}.

Note that by combining the sub-box method with the finite-size scaling formula (\ref{fss}) we can also obtain the values of the critical exponents.
To this aim we employ the data-collapse optimization method that has been previously used to extract critical exponents in various spin models~\cite{bhattacharjee2001measure, houdayer2004low}. To use this approach one first measures the variable $\mathcal{O}$ at different values of the control parameter and of the system size. All collected measurements form a set of $M$ data-points, where ${\mathcal{O}_i=\mathcal{O}(\Delta \mathcal{v}_i,L_i)}$ represent the $i$-th data point. If (\ref{fss}) holds, and the correct values of $\Delta \mathcal{v}_c$, $\zeta_\mathcal{O}$ and $\nu$ are known, the following quantity should be zero

\begin{equation}
    \mathcal{E}_\mathcal{O} = \sum_{i=1}^M \left [ {L_i}^{-\zeta_{\mathcal{O}/\nu}} \mathcal{O}(\Delta \mathcal{v}_{i},L_{i})-G_\mathcal{O}(\Delta \tilde{\mathcal{v}}_i) \right ]^2,
    \label{opti}
\end{equation}

\noindent where we have introduced the scaled variable ${\Delta \tilde{\mathcal{v}}_i = {L_i}^{1/\nu}(\Delta \mathcal{v}_{i}-\Delta \mathcal{v}_c)}$. Conversely, if we knew $G_\mathcal{O}$, we could minimize (\ref{opti}), with respect to $\Delta \mathcal{v}_c$, $\zeta_\mathcal{O}$ and $\nu$ to find their correct values. Although we generally do not know  $G_\mathcal{O}$, we can still approximate it from the $\mathcal{O}_i$ data. To this aim we divide the ${\Delta\tilde{\mathcal{v}}_i}$ values in a set of $W$ windows (with $M<W$) and average the values of the data-points ${{L_i}^{-\zeta_{\mathcal{O}/\nu}} \mathcal{O}_i}$ lying within the same window.
The value of $G_\mathcal{O}(\Delta \tilde{\mathcal{v}}_i)$ appearing in (\ref{opti}) can then be calculated by a linear interpolation of the window-averaged $G_\mathcal{O}$. With this technique the function $\mathcal{E}_\mathcal{O}$ can be then minimized to obtain the parameters $\zeta_\mathcal{O}$, $\nu$ and $\Delta \mathcal{v}_c$.

\section*{Results}

Before focusing on the critical behavior we start by discussing how simulations results compare with the mean-field picture explained above. We first numerically locate the boundaries of the MIPS-region, by performing a numerical scan in $\rho$ and $\Delta \mathcal{v}$ for the smallest system size ($N=3500$) distinguishing those state points in which phase-separation occurs from those in which the system is homogeneous. This can be quickly done by checking whether the density distribution (computed over all sub-boxes) has two detectable peaks. The resulting coarse phase-diagram is reported in Fig.~\ref{sktch}(b). Already from this data we can approximately locate the critical point to be at ${\Delta \mathcal{v}_c\approx20}$ and $\rho_c \approx 3$. Visual inspection of configurations immediately confirms that, upon increasing $\Delta \mathcal{v}$, QS active particles undergo a transition from a homogeneous state (Fig.~\ref{sktch}(e)) to phase separation (Fig.~\ref{sktch}(c)), with large fluctuations at intermediate $\Delta \mathcal{v}$-values (Fig.~\ref{sktch}(d)). The overall picture confirms that the microscopic model follows qualitatively the mean-field scenario discussed above, albeit with significant quantitative differences. In particular, we observe that the MIPS of the numerical model occurs at significantly higher $\Delta \mathcal{v}$ and $\rho$-values. Interestingly however, the shape of the numerical coexistence curve has a form reminescent of the theoretical spinodal line which is slightly asymmetric with the dense-fluid branch less steep than the gaseous one.
\begin{figure*}[t!]
\centering
\includegraphics[width=0.99\textwidth]{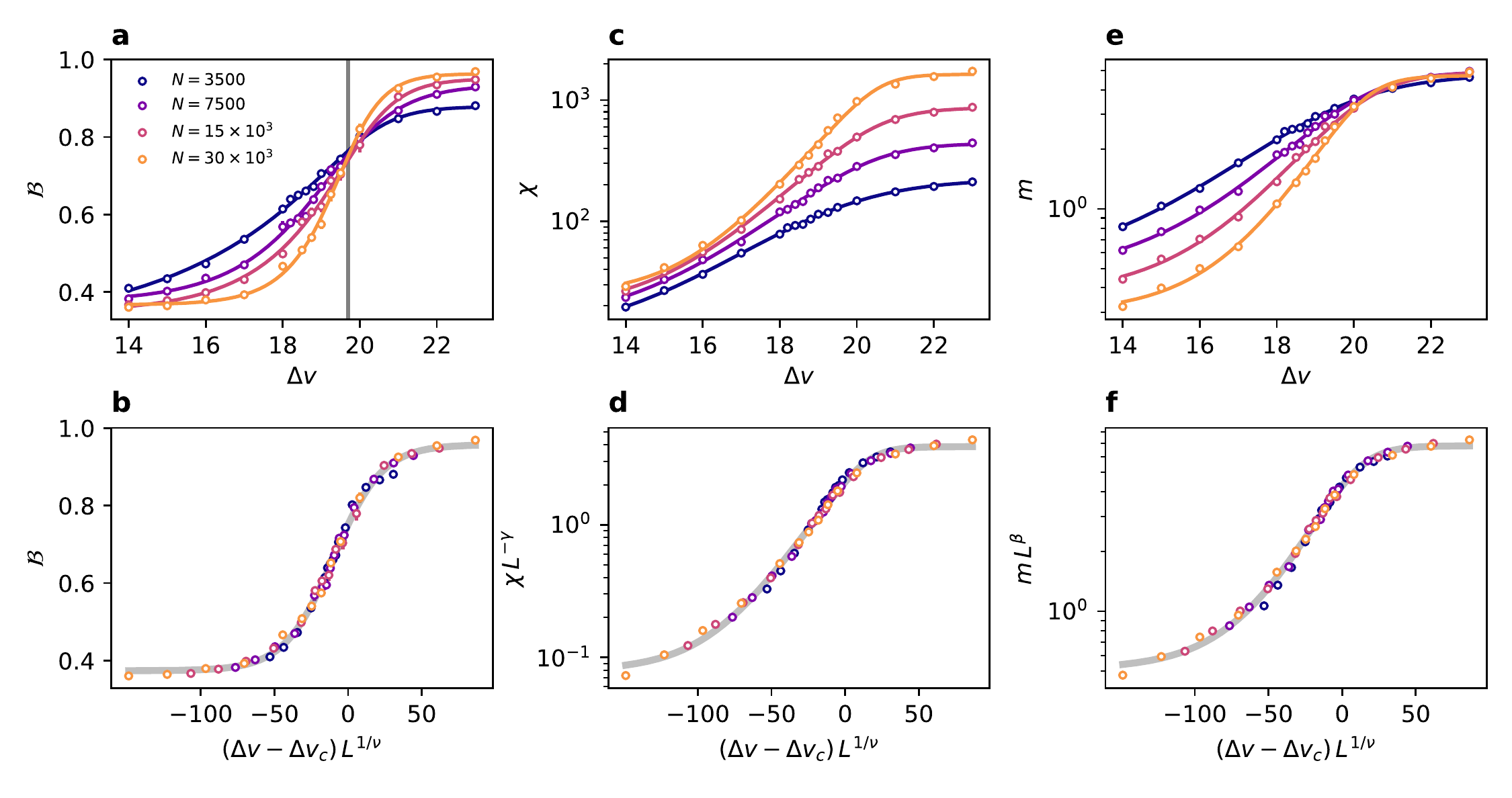}
\caption{
(\textbf{a}) Binder parameter $\mathcal{B}$ obtained in simulations (data points) as function of $\Delta \mathcal{v}$ for different system sizes (see legend). The full lines are fits with generalized logistic functions to guide the eye (also in (c) and (e)). The vertical line shows the value of the critical $\Delta \mathcal{v}$ which is close to the intersection of the curves.
(\textbf{b}) collapse of the data in (a) with the parameters $\nu$ and $\Delta \mathcal{v}_c$ determined by the optimization method. The full line is a fit of all data-points with the generalized logistic function as a guide to the eye (also in (d) and (f))
(\textbf{c}) susceptibility $\chi$ as a function of $\Delta \mathcal{v}$ for different system sizes.
(\textbf{d}) data of (c) collapsed with the critical exponent $\gamma$  
(\textbf{e}) order parameter $m$ vs $\Delta \mathcal{v}$ for different
system sizes.
(\textbf{f}) collapsed data of (e) with the exponent $\beta$.}
\label{scal_dens}
\end{figure*}

To precisely locate the critical point we proceed as in Ref.~\cite{maggi2021universality} to calculate the Binder parameter $\mathcal{B}$, introduced in the previous Section, as a function of $\rho$ at fixed $\Delta \mathcal{v}=18$ near the critical $\Delta \mathcal{v}$. According to Ref.~\cite{rovere1990gas}, $\mathcal{B}$ should exhibit a maximum at $\rho=\rho_c$ when plotted as a function of the density at fixed $\Delta \mathcal{v}$. We performed this analysis for the smallest system size and, fitting the data with a parabola, we get the critical density $\rho_c=2.821(0.041)$ (in the brackets we report the fit error). Since we are interested in the critical behavior of the system from now on we will fix $\rho=\rho_c$ for our analysis. To determine accurately the value of $\Delta \mathcal{v}_c$ and of the exponent $\nu$ we first compute $\mathcal{B}$ at various values of $\Delta \mathcal{v}$ and $L$ (simulation data are shown in Fig.~\ref{scal_dens}(a)) and we minimize (\ref{opti}) with respect to $\nu$ and $\Delta \mathcal{v}_c$ (which are considered as free fitting parameters). The optimization yields $\Delta \mathcal{v}_{c}=19.695(0.048)$ and $\nu=1.040(0.045)$. We note that the exponents $\nu$ is close to the Ising value and that, when we use these parameters to scale data, we obtain a good collapse as shown in Fig.~\ref{scal_dens}(b). Note also that, the the obtained $\Delta \mathcal{v}_c$ 
corresponds, to a good approximation, with value at which the Binder cumulants cross, as shown by the vertical line in Fig.~\ref{scal_dens}(a)

\begin{figure}
 \centering
\includegraphics[width=0.99\columnwidth]{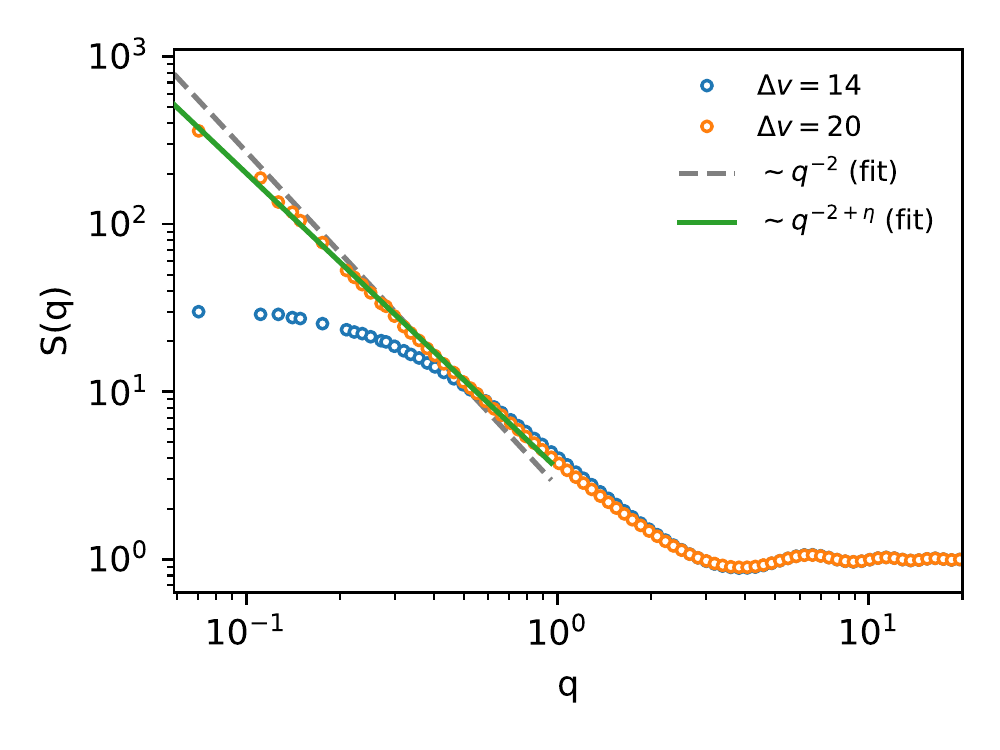}
\caption{Static structure factor $S(q)$ computed  for the
largest system (${N = 30\times10^3}$) at two state points with the same $\rho=\rho_c$ but different $\Delta \mathcal{v}$ (see legend).  $S(q)$ tends to a constant at low $q$ in the homogeneous phase (blue points). Close to the critical point (orange data-points) the structure factor is well fitted, for small $q$ values, by a power law $S(q)\sim q^{-2+\eta}$ (full line). The dashed line represents the fit with $S(q)\sim q^{-2}$ (mean-field).
\label{fig:sq}
}
\end{figure}

\begin{figure*}
\centering
\includegraphics[width=0.99\textwidth]{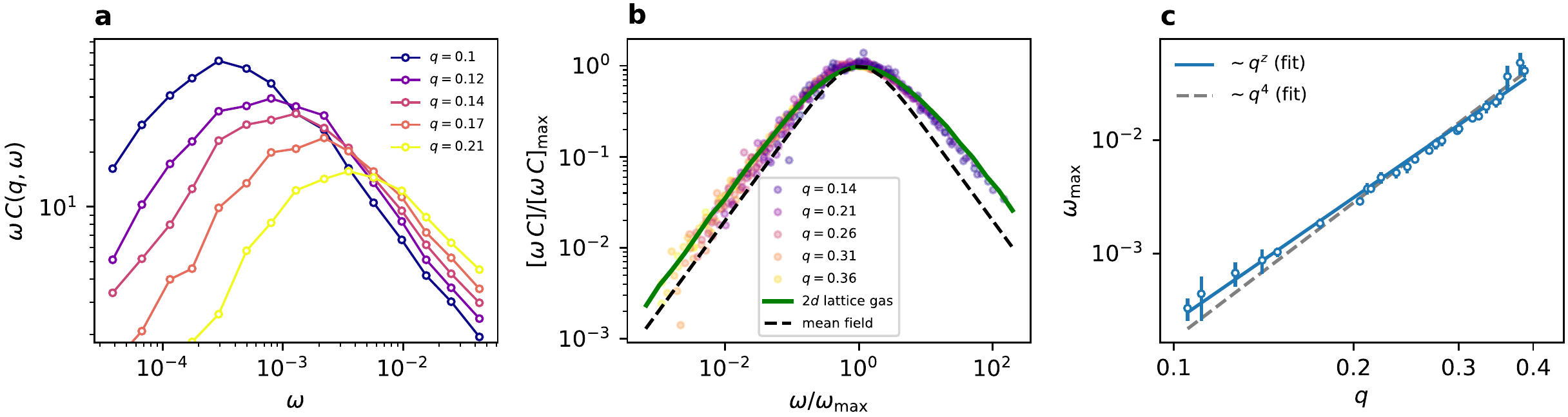}
\caption{
(\textbf{a}) Frequency-resolved correlation functions at various q values (see legend). 
(\textbf{b}) Scaling of the data in (a) by their amplitude and peak position. The green curve represents the scaled correlator from the equilibrium lattice gas. The dashed curve is the correlator from the Gaussian-field theory shown for comparison. 
(\textbf{c}) Peak frequency $\omega_{\mathrm{max}}$, from data in (a), plotted as a function of $q$ (the error-bar on each data-point corresponds to the error made in fitting the peak position). The full line is a power-law fit to estimate the critical exponent $z$. 
The dashed line is the fit with the fixed exponent $z=4$ (mean-field). }
\label{scal_dyn}
\end{figure*}
We now proceed to determine independently the exponents $\gamma$ and $\beta$.
To this aim, we fix the values of $\nu$ and $\Delta \mathcal{v}_c$ to those found previously, letting $\zeta_\mathcal{O}$ as a free parameter in the optimization. 
To estimate $\gamma$ we consider the susceptibility
${\chi=\langle (N_b - \langle N_b\rangle)^2\rangle/\langle N_b\rangle}$ at different $\Delta \mathcal{v}$ and $L$-values,
as shown in Fig.\ref{scal_dens} (c). We then minimize (\ref{opti}) with respect to $\zeta_\mathcal{O}=\gamma$ finding that $\gamma=1.763(0.070)$ (the error on $\nu$ is propagated), which is compatible with the Ising value. Data are well collapsed using this exponent as shown in Fig.~\ref{scal_dens}(d). For estimating $\beta$ we consider the order parameter $m=\rho_{\mathrm{dens}}-\rho_{\mathrm{dil}}$, defined as the difference between the densities of the dense ($\rho_{\mathrm{dens}}$) and dilute ($\rho_{\mathrm{dil}}$) phases~\cite{siebert2018critical}. The values of $m$, collected at several $\Delta \mathcal{v}$ and $L$, are shown in Fig.~\ref{scal_dens}(e). The minimization of the error-function (\ref{opti}) with $\zeta_\mathcal{O}=-\beta$ gives ${\beta=0.116(0.013)}$ (which is close to the Ising $\beta=0.125$) and a scaling of data with this $\beta$ provides a good collapse as shown Fig.~\ref{scal_dens}(f).
To conclude our analysis of the static critical exponents we now estimate the exponent $\eta$ which controls the behavior of the critical static structure factor $S(q)$ at low wave-vectors~$q$. 
We consider $S(q)=\mathcal{A}\,\langle \rho_\mathbf{-q}\rho_\mathbf{q}\rangle$, where $\rho_\mathbf{q}$ is the real part of the Fourier transform of the density field at time $t$: ${\rho_\mathbf{q}(t) = \sum_{i=1}^N \cos(\mathbf{q}\cdot \mathbf{x}_i(t))}$, where ${\mathbf{q}=2 \pi\, (n_x/L_x,n_y/L_y)}$ represents the wave-vector 
($n_{x,y}=\mathbb{Z}$). Here we assume that rotational and  time-translational invariance hold so that $S(q)$ is averaged over all $\mathbf{q}$ with the same modulus $q=|\mathbf{q}|$ and over all times $t$.
Moreover, the prefactor $\mathcal{A}$ is chosen such that $S(q\rightarrow\infty)=1$.
The $S(q)$, calculated for the largest system size, is shown in Fig.~\ref{fig:sq} for two different $\Delta \mathcal{v}$-values across the transition. While far from the critical point the $S(q)$ flattens at low $q$, close to $\Delta \mathcal{v}_{c}$ the structure factor exhibits a power-law behavior. By fitting the low-$q$ data with $S(q)\sim q^{-2+\eta}$ we obtain $2-\eta=1.765(0.012)$, which is close to the Ising value ($2-\eta=1.75$). Note that the slope of the fit is appreciably different from the typical mean-field decay $S(q)\sim q^{-2}$ which is also reported Fig.~\ref{fig:sq}.

Finally, we focus on the critical dynamics of the QS system. As in Ref.~\cite{MaggiCommPhys2022}, we compute the time auto-correlation function ${C(q,t)=2N^{-1}\langle \rho_{\mathbf{-q}}(t+s)\rho_{\mathbf{q}}(s)\rangle}$ to  characterizes the dynamics of spontaneous fluctuations at different time and length-scales.
The $C(q,t)$ is computed for various $q$-s and at the near-critical value $\Delta \mathcal{v}=20$.
From correlations we compute the spectra ${C(q, \omega)=\int dt e^{i\omega\,t} C(q, t)}$. The functions ${\omega\,C(q, \omega)}$ are shown in Fig.~\ref{scal_dyn}(a).
We see that ${\omega\,C(q, \omega)}$ grows in amplitude upon lowering $q$ and its peak shifts towards lower frequencies revealing the characteristic slowing-down of the system's dynamics at large length-scales. 
As discussed in Ref.~\cite{MaggiCommPhys2022} the value of $\omega$ at which ${\omega\,C(q, \omega)}$ reaches its maximum (denoted as $\omega_\mathrm{max}$) can be associated with the system's relaxation frequency at that $q$. Moreover in Ref.~\cite{MaggiCommPhys2022} it was shown that AOUPs may exhibit deviations from the scaling regime at frequencies as high as the relaxation rate of the active force $\tau^{-1}$ (appearing in the equations of motions (\ref{vx}) and (\ref{vx2})). 
For this reason we restrict our analysis to $q$-values that are low enough so that $\omega_\mathrm{max} \ll 2\pi/\tau$. 
\begin{table}[b]
\caption{Critical exponents estimated from simulations of the two-dimensional QS active system. The exponents of the  Ising model with conserved magnetization are reported for comparison.}
\label{tab:table1}
\begin{tabular}{cccc}
\textrm{Exponent}&
\textrm{Ising value}&
\textrm{fitted value}&
\textrm{error}\\
$\nu$ & 1 & 1.040 & 0.045\\
$\gamma$ & 1.75 & 1.763 & 0.070\\
$\beta$ & 0.125 & 0.116 & 0.013\\
$\eta$ & 0.25 & 0.235 & 0.011\\
$z$    & 3.75  & 3.640 & 0.073\\
\end{tabular}
\end{table}
Within this $q$-range we first characterize the shape of ${\omega\,C(q,\omega)}$.
This is an interesting feature to consider since it is well know~\cite{tauber2014critical} that, at the critical point, the relaxation spectrum $C(q,\omega)$ could be rewritten in terms of a universal scaling function $\mathcal{C}$ as follows:

\begin{equation}
\label{dysca}
\omega \, C(q,\omega) = 
  q^{2 z+2-\eta} \, 
\mathcal{C}(\omega/q^{z})
\end{equation}

\noindent Eq.~(\ref{dysca}) implies that $\omega_\mathrm{max}\sim q^z$ and that all ${\omega\,C(q,\omega)}$ should have the same shape described by $\mathcal{C}$. In Fig.~\ref{scal_dyn}(b) we check this by plotting data of Fig.~\ref{scal_dyn}(a) rescaled by their maxima, finding a good superposition.
Moreover, since $\mathcal{C}$ should be a universal scaling function we plot on top of our data the scaled ${\omega\,C(q,\omega)}$ of the critical two-dimensional lattice gas in equilibrium (see Ref.s~\cite{MaggiCommPhys2022,caprini2020activity} for details). We find that the data of the QS active system follow quite well the correlator of the lattice gas simulations (especially at large $\omega/\omega_\mathrm{max}$). Data are also compared with the result of the dynamical Gaussian  field-theory~\cite{tauber2014critical} which yields ${\omega\,C(q,\omega) \sim 
\omega/(\omega^2+{\omega_\mathrm{max}^2})}$. Fig.~\ref{scal_dyn}(b) shows that this formula (which should be valid only for $d>4$) sensibly deviates from the data-points as it should. 
As mentioned above we expect that the peak position follows $\omega_\mathrm{max}\sim q^z$ and this allows us to estimate the dynamic critical exponent $z$.  In Fig.~\ref{scal_dyn}(c) we show $\omega_\mathrm{max}$ as a function of $q$ and the direct fit with a power-law giving $z=3.640(0.073)$ which is also compatible with $z=3.75$ of the Ising model, with conserved magnetization, in $d=2$.

All exponents estimated in the present work are summarized in Table~1.

\section*{Conclusions}
In this work we have investigated the critical behavior of an active system interacting via QS.
In the model each particle \emph{senses} the number of neighbours, within a given cut-off distance, and then varies its speed according to a specific rule. 
The mean-field theory suggests that this model has a spinodal line ending in a critical point. 
To check the validity of the theoretical picture we have performed large-scale numerical simulations on GPU, finding that system fully does phase-separates and that the coexistence region terminates in a  motility induced critical point.
To address the problem of the universality class, we use finite-size scaling analysis to measure static and dynamic critical exponents. We have found that these exponents are in substantial agreement with those of the two-dimensional Ising model with conserved magnetization. 

Our study opens different possibilities for future investigations on critical active systems. From  a more general
perspective it would be interesting to understand if and how the QS interaction rules could be changed to destabilize
the critical point and to possibly observe micro-phase separation as suggested by recent active
field-theories~\cite{caballero2018bulk}. Moreover it would be interesting to understand if and how the fluctuation dissipation theorem could be violated by critical QS particles. In particular, it could be interesting to understand whether the breakdown of the theorem differs from the one observed for purely repulsive AOUPs~\cite{MaggiCommPhys2022}.





\bibliography{mpbib}

\providecommand*{\mcitethebibliography}{\thebibliography}
\csname @ifundefined\endcsname{endmcitethebibliography}
{\let\endmcitethebibliography\endthebibliography}{}
\begin{mcitethebibliography}{39}
\providecommand*{\natexlab}[1]{#1}
\providecommand*{\mciteSetBstSublistMode}[1]{}
\providecommand*{\mciteSetBstMaxWidthForm}[2]{}
\providecommand*{\mciteBstWouldAddEndPuncttrue}
  {\def\EndOfBibitem{\unskip.}}
\providecommand*{\mciteBstWouldAddEndPunctfalse}
  {\let\EndOfBibitem\relax}
\providecommand*{\mciteSetBstMidEndSepPunct}[3]{}
\providecommand*{\mciteSetBstSublistLabelBeginEnd}[3]{}
\providecommand*{\EndOfBibitem}{}
\mciteSetBstSublistMode{f}
\mciteSetBstMaxWidthForm{subitem}
{(\emph{\alph{mcitesubitemcount}})}
\mciteSetBstSublistLabelBeginEnd{\mcitemaxwidthsubitemform\space}
{\relax}{\relax}

\bibitem[Miller and Bassler(2001)]{MillerAnnualRev2001}
M.~B. Miller and B.~L. Bassler, \emph{Annu. Rev. Microbiol.}, 2001,
  \textbf{55}, 165--199\relax
\mciteBstWouldAddEndPuncttrue
\mciteSetBstMidEndSepPunct{\mcitedefaultmidpunct}
{\mcitedefaultendpunct}{\mcitedefaultseppunct}\relax
\EndOfBibitem
\bibitem[Nealson \emph{et~al.}(1970)Nealson, Platt, and Hastings]{Nealson1970}
K.~H. Nealson, T.~Platt and J.~W. Hastings, \emph{J. Bacteriol.}, 1970,
  \textbf{104}, 313--322\relax
\mciteBstWouldAddEndPuncttrue
\mciteSetBstMidEndSepPunct{\mcitedefaultmidpunct}
{\mcitedefaultendpunct}{\mcitedefaultseppunct}\relax
\EndOfBibitem
\bibitem[Zhu \emph{et~al.}(2002)Zhu, Miller, Vance, Dziejman, B.L., and
  J.J.]{ZhuPNAS2002}
J.~Zhu, M.~Miller, R.~Vance, M.~Dziejman, B.~B.L. and M.~J.J., \emph{Proc.
  Natl. Acad. Sci. USA}, 2002, \textbf{99(5)}, 3129--3134\relax
\mciteBstWouldAddEndPuncttrue
\mciteSetBstMidEndSepPunct{\mcitedefaultmidpunct}
{\mcitedefaultendpunct}{\mcitedefaultseppunct}\relax
\EndOfBibitem
\bibitem[Hammer and Bassler(2003)]{Hammer2003}
B.~Hammer and B.~Bassler, \emph{Mol Microbiol.}, 2003, \textbf{50(1)},
  101--104\relax
\mciteBstWouldAddEndPuncttrue
\mciteSetBstMidEndSepPunct{\mcitedefaultmidpunct}
{\mcitedefaultendpunct}{\mcitedefaultseppunct}\relax
\EndOfBibitem
\bibitem[Daniels \emph{et~al.}(2004)Daniels, Vanderleyden, and
  Michiels]{Daniels2004}
R.~Daniels, J.~Vanderleyden and J.~Michiels, \emph{FEMS Microbiol. Rev.}, 2004,
  \textbf{28}, 261--289\relax
\mciteBstWouldAddEndPuncttrue
\mciteSetBstMidEndSepPunct{\mcitedefaultmidpunct}
{\mcitedefaultendpunct}{\mcitedefaultseppunct}\relax
\EndOfBibitem
\bibitem[Gomez-Solano \emph{et~al.}(2017)Gomez-Solano, Samin, Lozano,
  Ruedas-Batuecas, van Roij, and Bechinger]{Gomez-SolanoSciRep2017}
J.~R. Gomez-Solano, S.~Samin, C.~Lozano, P.~Ruedas-Batuecas, R.~van Roij and
  C.~Bechinger, \emph{Sci. Rep.}, 2017, \textbf{7}, 14891\relax
\mciteBstWouldAddEndPuncttrue
\mciteSetBstMidEndSepPunct{\mcitedefaultmidpunct}
{\mcitedefaultendpunct}{\mcitedefaultseppunct}\relax
\EndOfBibitem
\bibitem[Jiang \emph{et~al.}(2010)Jiang, Yoshinaga, and
  Sano]{PhysRevLett.105.268302}
H.-R. Jiang, N.~Yoshinaga and M.~Sano, \emph{Phys. Rev. Lett.}, 2010,
  \textbf{105}, 268302\relax
\mciteBstWouldAddEndPuncttrue
\mciteSetBstMidEndSepPunct{\mcitedefaultmidpunct}
{\mcitedefaultendpunct}{\mcitedefaultseppunct}\relax
\EndOfBibitem
\bibitem[Maggi \emph{et~al.}(2016)Maggi, Simmchen, Saglimbeni, Katuri, Dipalo,
  De~Angelis, Sanchez, and Di~Leonardo]{maggi2016self}
C.~Maggi, J.~Simmchen, F.~Saglimbeni, J.~Katuri, M.~Dipalo, F.~De~Angelis,
  S.~Sanchez and R.~Di~Leonardo, \emph{Small}, 2016, \textbf{12},
  446--451\relax
\mciteBstWouldAddEndPuncttrue
\mciteSetBstMidEndSepPunct{\mcitedefaultmidpunct}
{\mcitedefaultendpunct}{\mcitedefaultseppunct}\relax
\EndOfBibitem
\bibitem[B{\"a}uerle \emph{et~al.}(2018)B{\"a}uerle, Fischer, Speck, and
  Bechinger]{SpeckNatComm2018}
T.~B{\"a}uerle, A.~Fischer, T.~Speck and C.~Bechinger, \emph{Nat. Commun.},
  2018, \textbf{9}, 1--8\relax
\mciteBstWouldAddEndPuncttrue
\mciteSetBstMidEndSepPunct{\mcitedefaultmidpunct}
{\mcitedefaultendpunct}{\mcitedefaultseppunct}\relax
\EndOfBibitem
\bibitem[Lavergne \emph{et~al.}(2019)Lavergne, Wendehenne, and
  B\"{a}uerle~T]{Lavergne2019}
F.~Lavergne, H.~Wendehenne and B.~C. B\"{a}uerle~T, \emph{Science}, 2019,
  \textbf{364(6435)}, 70--74\relax
\mciteBstWouldAddEndPuncttrue
\mciteSetBstMidEndSepPunct{\mcitedefaultmidpunct}
{\mcitedefaultendpunct}{\mcitedefaultseppunct}\relax
\EndOfBibitem
\bibitem[Cates and Tailleur(2015)]{CatesAnnRev2015}
M.~E. Cates and J.~Tailleur, \emph{Annu. Rev. Condens. Matter Phys.}, 2015,
  \textbf{6}, 219--244\relax
\mciteBstWouldAddEndPuncttrue
\mciteSetBstMidEndSepPunct{\mcitedefaultmidpunct}
{\mcitedefaultendpunct}{\mcitedefaultseppunct}\relax
\EndOfBibitem
\bibitem[Solon \emph{et~al.}(2018)Solon, Stenhammar, Cates, Kafri, and
  Tailleur]{SolonPRE2018}
A.~P. Solon, J.~Stenhammar, M.~E. Cates, Y.~Kafri and J.~Tailleur, \emph{Phys.
  Rev. E}, 2018, \textbf{97}, 020602\relax
\mciteBstWouldAddEndPuncttrue
\mciteSetBstMidEndSepPunct{\mcitedefaultmidpunct}
{\mcitedefaultendpunct}{\mcitedefaultseppunct}\relax
\EndOfBibitem
\bibitem[Maggi \emph{et~al.}(2021)Maggi, Paoluzzi, Crisanti, Zaccarelli, and
  Gnan]{maggi2021universality}
C.~Maggi, M.~Paoluzzi, A.~Crisanti, E.~Zaccarelli and N.~Gnan, \emph{Soft
  Matter}, 2021, \textbf{17}, 3807--3812\relax
\mciteBstWouldAddEndPuncttrue
\mciteSetBstMidEndSepPunct{\mcitedefaultmidpunct}
{\mcitedefaultendpunct}{\mcitedefaultseppunct}\relax
\EndOfBibitem
\bibitem[Maggi \emph{et~al.}(2022)Maggi, Gnan, Paoluzzi, Zaccarelli, and
  Crisanti]{MaggiCommPhys2022}
C.~Maggi, N.~Gnan, M.~Paoluzzi, E.~Zaccarelli and A.~Crisanti, \emph{Commun.
  Phys.}, 2022, \textbf{5}, 55\relax
\mciteBstWouldAddEndPuncttrue
\mciteSetBstMidEndSepPunct{\mcitedefaultmidpunct}
{\mcitedefaultendpunct}{\mcitedefaultseppunct}\relax
\EndOfBibitem
\bibitem[Partridge and Lee(2019)]{PhysRevLett.123.068002}
B.~Partridge and C.~F. Lee, \emph{Phys. Rev. Lett.}, 2019, \textbf{123},
  068002\relax
\mciteBstWouldAddEndPuncttrue
\mciteSetBstMidEndSepPunct{\mcitedefaultmidpunct}
{\mcitedefaultendpunct}{\mcitedefaultseppunct}\relax
\EndOfBibitem
\bibitem[Siebert \emph{et~al.}(2018)Siebert, Dittrich, Schmid, Binder, Speck,
  and Virnau]{siebert2018critical}
J.~T. Siebert, F.~Dittrich, F.~Schmid, K.~Binder, T.~Speck and P.~Virnau,
  \emph{Phys. Rev. E}, 2018, \textbf{98}, 030601\relax
\mciteBstWouldAddEndPuncttrue
\mciteSetBstMidEndSepPunct{\mcitedefaultmidpunct}
{\mcitedefaultendpunct}{\mcitedefaultseppunct}\relax
\EndOfBibitem
\bibitem[Dittrich \emph{et~al.}(2021)Dittrich, Speck, and
  Virnau]{dittrich2021critical}
F.~Dittrich, T.~Speck and P.~Virnau, \emph{Eur. Phys. J. E}, 2021, \textbf{44},
  1--10\relax
\mciteBstWouldAddEndPuncttrue
\mciteSetBstMidEndSepPunct{\mcitedefaultmidpunct}
{\mcitedefaultendpunct}{\mcitedefaultseppunct}\relax
\EndOfBibitem
\bibitem[Caporusso \emph{et~al.}(2020)Caporusso, Digregorio, Levis,
  Cugliandolo, and Gonnella]{PhysRevLett.125.178004}
C.~B. Caporusso, P.~Digregorio, D.~Levis, L.~F. Cugliandolo and G.~Gonnella,
  \emph{Phys. Rev. Lett.}, 2020, \textbf{125}, 178004\relax
\mciteBstWouldAddEndPuncttrue
\mciteSetBstMidEndSepPunct{\mcitedefaultmidpunct}
{\mcitedefaultendpunct}{\mcitedefaultseppunct}\relax
\EndOfBibitem
\bibitem[Caprini \emph{et~al.}(2020)Caprini, Marconi, Maggi, Paoluzzi, and
  Puglisi]{caprini2020hidden}
L.~Caprini, U.~M.~B. Marconi, C.~Maggi, M.~Paoluzzi and A.~Puglisi, \emph{Phys.
  Rev. Research}, 2020, \textbf{2}, 023321\relax
\mciteBstWouldAddEndPuncttrue
\mciteSetBstMidEndSepPunct{\mcitedefaultmidpunct}
{\mcitedefaultendpunct}{\mcitedefaultseppunct}\relax
\EndOfBibitem
\bibitem[Marconi \emph{et~al.}(2016)Marconi, Paoluzzi, and Maggi]{Marconi3}
U.~M.~B. Marconi, M.~Paoluzzi and C.~Maggi, \emph{Molecular Physics}, 2016,
  \textbf{114}, 2400--2410\relax
\mciteBstWouldAddEndPuncttrue
\mciteSetBstMidEndSepPunct{\mcitedefaultmidpunct}
{\mcitedefaultendpunct}{\mcitedefaultseppunct}\relax
\EndOfBibitem
\bibitem[Fodor \emph{et~al.}(2016)Fodor, Nardini, Cates, Tailleur, Visco, and
  van Wijland]{Fodor16}
E.~Fodor, C.~Nardini, M.~E. Cates, J.~Tailleur, P.~Visco and F.~van Wijland,
  \emph{Phys. Rev. Lett.}, 2016, \textbf{117}, 038103\relax
\mciteBstWouldAddEndPuncttrue
\mciteSetBstMidEndSepPunct{\mcitedefaultmidpunct}
{\mcitedefaultendpunct}{\mcitedefaultseppunct}\relax
\EndOfBibitem
\bibitem[Marini Bettolo~Marconi \emph{et~al.}(2017)Marini Bettolo~Marconi,
  Maggi, and Paoluzzi]{Marconi17}
U.~Marini Bettolo~Marconi, C.~Maggi and M.~Paoluzzi, \emph{The Journal of
  chemical physics}, 2017, \textbf{147}, 024903\relax
\mciteBstWouldAddEndPuncttrue
\mciteSetBstMidEndSepPunct{\mcitedefaultmidpunct}
{\mcitedefaultendpunct}{\mcitedefaultseppunct}\relax
\EndOfBibitem
\bibitem[Dal~Cengio \emph{et~al.}(2019)Dal~Cengio, Levis, and
  Pagonabarraga]{dal2019linear}
S.~Dal~Cengio, D.~Levis and I.~Pagonabarraga, \emph{Physical Review Letters},
  2019, \textbf{123}, 238003\relax
\mciteBstWouldAddEndPuncttrue
\mciteSetBstMidEndSepPunct{\mcitedefaultmidpunct}
{\mcitedefaultendpunct}{\mcitedefaultseppunct}\relax
\EndOfBibitem
\bibitem[Martin \emph{et~al.}(2021)Martin, O'Byrne, Cates, Fodor, Nardini,
  Tailleur, and van Wijland]{martin2021statistical}
D.~Martin, J.~O'Byrne, M.~E. Cates, {\'E}.~Fodor, C.~Nardini, J.~Tailleur and
  F.~van Wijland, \emph{Physical Review E}, 2021, \textbf{103}, 032607\relax
\mciteBstWouldAddEndPuncttrue
\mciteSetBstMidEndSepPunct{\mcitedefaultmidpunct}
{\mcitedefaultendpunct}{\mcitedefaultseppunct}\relax
\EndOfBibitem
\bibitem[Risken(1996)]{risken1996fokker}
H.~Risken, \emph{The Fokker-Planck Equation}, Springer, 1996, pp. 63--95\relax
\mciteBstWouldAddEndPuncttrue
\mciteSetBstMidEndSepPunct{\mcitedefaultmidpunct}
{\mcitedefaultendpunct}{\mcitedefaultseppunct}\relax
\EndOfBibitem
\bibitem[Gardiner
  \emph{et~al.}(1985)Gardiner\emph{et~al.}]{gardiner1985handbook}
C.~W. Gardiner \emph{et~al.}, \emph{Handbook of stochastic methods}, springer
  Berlin, 1985, vol.~3\relax
\mciteBstWouldAddEndPuncttrue
\mciteSetBstMidEndSepPunct{\mcitedefaultmidpunct}
{\mcitedefaultendpunct}{\mcitedefaultseppunct}\relax
\EndOfBibitem
\bibitem[H{\"a}nggi and Jung(1995)]{Hanggi95}
P.~H{\"a}nggi and P.~Jung, \emph{Adv. Chem. Phys.}, 1995, \textbf{89},
  239--326\relax
\mciteBstWouldAddEndPuncttrue
\mciteSetBstMidEndSepPunct{\mcitedefaultmidpunct}
{\mcitedefaultendpunct}{\mcitedefaultseppunct}\relax
\EndOfBibitem
\bibitem[van Kampen(1964)]{vanKampen}
N.~G. van Kampen, \emph{Phys. Rev.}, 1964, \textbf{135}, A362--A369\relax
\mciteBstWouldAddEndPuncttrue
\mciteSetBstMidEndSepPunct{\mcitedefaultmidpunct}
{\mcitedefaultendpunct}{\mcitedefaultseppunct}\relax
\EndOfBibitem
\bibitem[Tailleur and Cates(2008)]{Tailleur08}
J.~Tailleur and M.~E. Cates, \emph{Phys. Rev. Lett.}, 2008, \textbf{100},
  218103\relax
\mciteBstWouldAddEndPuncttrue
\mciteSetBstMidEndSepPunct{\mcitedefaultmidpunct}
{\mcitedefaultendpunct}{\mcitedefaultseppunct}\relax
\EndOfBibitem
\bibitem[Cates and Tailleur(2015)]{cates2015motility}
M.~E. Cates and J.~Tailleur, \emph{Annu. Rev. Condens. Matter Phys.}, 2015,
  \textbf{6}, 219--244\relax
\mciteBstWouldAddEndPuncttrue
\mciteSetBstMidEndSepPunct{\mcitedefaultmidpunct}
{\mcitedefaultendpunct}{\mcitedefaultseppunct}\relax
\EndOfBibitem
\bibitem[Binder(1981)]{binder1981finite}
K.~Binder, \emph{Zeitschrift f{\"u}r Physik B Condensed Matter}, 1981,
  \textbf{43}, 119--140\relax
\mciteBstWouldAddEndPuncttrue
\mciteSetBstMidEndSepPunct{\mcitedefaultmidpunct}
{\mcitedefaultendpunct}{\mcitedefaultseppunct}\relax
\EndOfBibitem
\bibitem[Rovere \emph{et~al.}(1988)Rovere, Hermann, and
  Binder]{rovere1988block}
M.~Rovere, D.~Hermann and K.~Binder, \emph{EPL (Europhysics Letters)}, 1988,
  \textbf{6}, 585\relax
\mciteBstWouldAddEndPuncttrue
\mciteSetBstMidEndSepPunct{\mcitedefaultmidpunct}
{\mcitedefaultendpunct}{\mcitedefaultseppunct}\relax
\EndOfBibitem
\bibitem[Rovere \emph{et~al.}(1990)Rovere, Heermann, and Binder]{rovere1990gas}
M.~Rovere, D.~W. Heermann and K.~Binder, \emph{J. Condens. Matter Phys.}, 1990,
  \textbf{2}, 7009\relax
\mciteBstWouldAddEndPuncttrue
\mciteSetBstMidEndSepPunct{\mcitedefaultmidpunct}
{\mcitedefaultendpunct}{\mcitedefaultseppunct}\relax
\EndOfBibitem
\bibitem[Rovere \emph{et~al.}(1993)Rovere, Nielaba, and
  Binder]{rovere1993simulation}
M.~Rovere, P.~Nielaba and K.~Binder, \emph{Zeitschrift f{\"u}r Physik B
  Condensed Matter}, 1993, \textbf{90}, 215--228\relax
\mciteBstWouldAddEndPuncttrue
\mciteSetBstMidEndSepPunct{\mcitedefaultmidpunct}
{\mcitedefaultendpunct}{\mcitedefaultseppunct}\relax
\EndOfBibitem
\bibitem[Bhattacharjee and Seno(2001)]{bhattacharjee2001measure}
S.~M. Bhattacharjee and F.~Seno, \emph{Journal of Physics A: Mathematical and
  General}, 2001, \textbf{34}, 6375\relax
\mciteBstWouldAddEndPuncttrue
\mciteSetBstMidEndSepPunct{\mcitedefaultmidpunct}
{\mcitedefaultendpunct}{\mcitedefaultseppunct}\relax
\EndOfBibitem
\bibitem[Houdayer and Hartmann(2004)]{houdayer2004low}
J.~Houdayer and A.~K. Hartmann, \emph{Physical Review B}, 2004, \textbf{70},
  014418\relax
\mciteBstWouldAddEndPuncttrue
\mciteSetBstMidEndSepPunct{\mcitedefaultmidpunct}
{\mcitedefaultendpunct}{\mcitedefaultseppunct}\relax
\EndOfBibitem
\bibitem[T{\"a}uber(2014)]{tauber2014critical}
U.~C. T{\"a}uber, \emph{Critical dynamics: a field theory approach to
  equilibrium and non-equilibrium scaling behavior}, Cambridge University
  Press, 2014\relax
\mciteBstWouldAddEndPuncttrue
\mciteSetBstMidEndSepPunct{\mcitedefaultmidpunct}
{\mcitedefaultendpunct}{\mcitedefaultseppunct}\relax
\EndOfBibitem
\bibitem[Caprini \emph{et~al.}(2020)Caprini, Cecconi, Maggi, and
  Marconi]{caprini2020activity}
L.~Caprini, F.~Cecconi, C.~Maggi and U.~M.~B. Marconi, \emph{Physical Review
  Research}, 2020, \textbf{2}, 043359\relax
\mciteBstWouldAddEndPuncttrue
\mciteSetBstMidEndSepPunct{\mcitedefaultmidpunct}
{\mcitedefaultendpunct}{\mcitedefaultseppunct}\relax
\EndOfBibitem
\bibitem[Caballero \emph{et~al.}(2018)Caballero, Nardini, and
  Cates]{caballero2018bulk}
F.~Caballero, C.~Nardini and M.~E. Cates, \emph{Journal of Statistical
  Mechanics: Theory and Experiment}, 2018, \textbf{2018}, 123208\relax
\mciteBstWouldAddEndPuncttrue
\mciteSetBstMidEndSepPunct{\mcitedefaultmidpunct}
{\mcitedefaultendpunct}{\mcitedefaultseppunct}\relax
\EndOfBibitem
\end{mcitethebibliography}
\bibliographystyle{rsc}
\end{document}